# Gradient-index Solar Sail and its Optimal Orbital Control

Shahin Firuzi, Yu Song, Shengping Gong[*]

Tsinghua University, School of Aerospace Engineering, Beijing, 100084, People's Republic of China
[*]gongsp@tsinghua.edu.cn

**ABSTRACT**

Solar sails with the capability of generating a tangential radiation pressure at the sun-pointing attitude, such as refractive sails can provide more efficient methods for attitude and orbital control of sailcraft. This paper presents the concept of gradient-index sail as an advanced class of refractive sail, which operates by guiding the solar radiation through a structure made of graded refractive index material. The design of the sail's refractive index distribution is performed by transformation optics, and the resultant index realized by the effective refractive index of non-resonant bulk metamaterials made of silica. The performance of the sail was evaluated by using ray tracing for a broad spectrum of solar radiation under the normal incidence angle, which showed an efficiency of 90.5% for generation of a tangential radiation pressure. We also studied the orbital control of the tangential-radiation-pressure-generating sails, and showed that the full orbital control, including the modification of orbital axes, eccentricity, and inclination can be applied by changing the attitude of the sail merely around the sun-sail axis, while the sail keeps the sun-pointing attitude at every point of the orbit.

## Introduction

Photonic propulsion can be divided into two main categories of propellant-less systems including solar sails and laser-driven sails, as well as the propellant-based ablative propulsion [1]. In the case of propellant-less propulsion, the sailcraft is pushed through space by the interaction with the electromagnetic (EM) radiation which can be the solar radiation or laser beam. Solar sailing which was firstly proposed in 1915 by Yakov Perelman and then presented scientifically by Konstantin Tsiolkowsky along with Fridrikh Tsander [2], has been demonstrated in several successful missions including IKAROS, NanoSail-D2, and LightSail. Solar sails have been also proposed for a large variety of missions including inner and outer solar system planetary and asteroid exploration, artificial Lagrange points, and displaced non-Keplerian orbits [3].

In different cases of propellant-less photonic propulsion, including solar sailing concepts [4, 5, 6, 7], and photonic devices [8, 9, 10], the radiation pressure is mostly generated due to the changes in the propagation direction of the EM waves. This change can be applied due to different interactions between the EM filed with matter, including reflection, refraction, and scattering. Solar sails conventionally operate by the reflection of the EM radiation from a highly reflective sheet, which results in the generation of a radiation pressure applied normal to the reflective surface. It has been shown that a radiation pressure tangential to the sail's surface can be also realized by the refraction of the solar radiation through a transparent sail composed of micro prisms [4]. The main characteristic of a refractive sail is the ability to generate a tangential radiation pressure at normal incidence angle, which as we will show in this paper, can be an advantage over the conventional sails which leads to simpler attitude and orbital control. In addition to the concept of refractive sail, a tangential radiation pressure can be also achieved through diffraction [5, 6, 11], and anomalous reflection of the solar radiation [7].

The maximum efficiency of a refractive sail at a normal incidence is computed to be about 40% of the maximum possible tangential radiation pressure. The maximum possible tangential radiation pressure at the normal incidence is equal to the radiation pressure generated due to the complete absorption, which in the Earth's orbit is equal to $P_{max} = 4.557 \times 10^{-6}$ Pa [12]. The decline in the magnitude of the tangential radiation pressure generated by a refractive sail is caused by two factors; the limitation in the maximum angle of the output beam due to the critical angle, and undesirable reflections on the sail's boundaries. These unpleasant features, however, can be dramatically improved by controlling the propagation of the EM filed through index-matched optical structures such as waveguides [13]. In optical devices such as optical fiber communication systems and lightwave circuits, the alteration of the direction of the EM waves has been traditionally performed by using optical waveguides. The invention of transformation optics (TO) in 2006 [14, 15], however, opened up a new avenue in designing the optical devices and enabled the precise control of the light flow by bending the space for light in a similar manner to the concept of curved space-time in the gravitational field [16]. TO approach has been used to design many remarkable optical devices such as invisibility cloaks, waveguides, wave shifters\splitters\combiners, flat lenses, illusion devices, highly directive antennas, et cetera [17]. Due to the ability of TO techniques in efficient guiding of the EM field, they may provide a suitable basis for designing refractive sails with a greater efficiency.

The orbital maneuver of solar sails can be performed by tuning the radiation pressure vector. In the case of conventional solar sails, since the radiation pressure is mostly normal to the sail's surface, the orbit and trajectory of the sails can be controlled through altering the attitude of the sail. It has been shown that the optimal inward and outward sun-centered orbit maneuvers with different inclinations can be achieved by altering the thrust vector through complicated attitude manipulations [18, 19, 20, 21], by using different mechanisms including gimbaled control boom, tip-mounted control vanes [22], reflectivity control devices [23], and moving masses [24], for the in-plane axes, as well as the refractive films for the normal axis [4]. In the case of tangential-radiation-pressure-generating sails such as refractive sail [4], diffractive sail [6], and the anomalous reflective sail [7], in a near-circular sun-centered orbit, the radiation pressure has a large component perpendicular to the sun-sail vector at a sun-pointing attitude. In this paper, for the first time, we will show that the tangential-radiation-pressure-generating sails can provide the possibility of orbital control through simple attitude maneuvers by using a combination of passive sun-pointing control [25], and an active control merely around the sail's normal axis which can be applied by using the refractive films [4].

In this paper, we study the possibility of designing a highly efficient refractive sail by using a specifically designed graded-index media by transformation optics method. The high efficiency of the graded-index sail can be described by the precise guidance of the light and the minimal reflection on the boundaries of the sail. We will also show how the heliocentric orbital maneuvers can be achieved for the tangential-radiation-pressure-generating solar sails. It will be shown that the control of the orbital radius and inclination can be realized by keeping a passive sun-pointing attitude and merely controlling the yaw angle (along the sun-sail axis), which is a great advantage over the reflective sails.

## Methods

### Design of the sail

The devices designed by TO can be implemented by metamaterials fabricated by using different materials including metals and dielectrics. However, metals have large losses at optical frequencies, where the solar radiation has the largest irradiance. Therefore, in order to obtain a low-loss design in the broad spectrum of solar radiation, an all-dielectric design is necessary. It has been shown that index-matched all dielectric designs can be achieved through conformal mapping (CM) technique [13]. The proposed gradient-index sail is composed of an array of identical micro-sized all-dielectric gradient-index waveguides which is shown in Fig. 1b. The structure of a single micro-waveguide along with the waveguide-fixed reference frame are shown in Fig. 1a, where the wide and narrow ports of the waveguide are the input and output ports, respectively. It can be seen that the waveguide consists of a bend and two straight guides at the input and output ports. The bend is constructed by two identical circular arcs of radius $r_w$ with displaced centers of distance $w_w$. As we will show later in this paper, the complete orbital control of the sail can be achieved by keeping a sun-pointing attitude. Therefore, the micro-waveguide can be designed merely for a constant normal incidence angle. The refractive index of the micro-waveguide is variable through the cross-section of the waveguide in order to guide the EM field in the desired manner, which is achieved by CM technique [13].

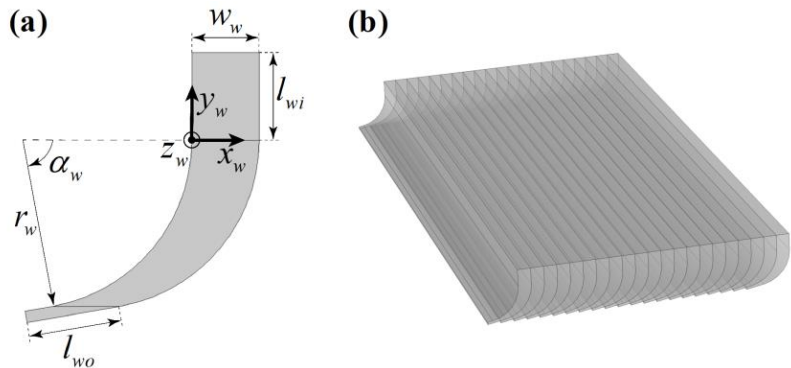

**Figure 1.** (a) Shows a single waveguide along with the waveguide-fixed reference frame. (b) Shows the gradient-index sail composed of an array of waveguides.

The conformal mapping is applied to transform the coordinates of the virtual domain to physical domain where the EM waves propagate in the straight and curved paths, respectively. The virtual and physical domains along with the coordinate systems and the boundary lines are shown in Fig. 2a and Fig. 2b, respectively. The conformal mapping can be applied by solving the Laplace's equation for $u$ and $v$, separately. However, for a strict CM, the mapping of either of $u$ and $v$, yields the same results. Therefore, the CM can be applied merely to one variable, which its results can be extended to the other variable. Here, without the loss of generality, CM is applied to $u$ and the refractive index distribution is derived accordingly. The Laplace's equation is applied to the physical domain for $u$ as

$$u_{xx}^2 + u_{yy}^2 = 0 \tag{1}$$

where the subscripts $xx$ and $yy$ correspond to the second-order partial derivatives with respect to $x$ and $y$, respectively. This equation is solved using the Dirichlet and Neumann boundary conditions, which is given as

$$u|_{\Gamma_1} = 0, u|_{\Gamma_3} = c_1 \quad \text{and} \quad N_{\Gamma_i} \cdot \nabla u|_{\Gamma_i} = 0 \text{ for } i = 2, 4 \tag{2}$$

where the nomenclature $N_{\Gamma i}$ is the normal vector to $\Gamma_i$, and $c_1$ is a design constant [13]. By having the solution of the Laplace's equation, the refractive index distribution of the waveguide can be given by

$$n_w = n_{vac} \sqrt{u_x^2 + u_y^2} \tag{3}$$

where $n_{vac}$ is the refractive index of the virtual domain, which by considering the sail to operate in vacuum, we have $n_{vac}=1$. In order to guarantee the resultant refractive index distribution to be realized by using dielectric materials, the minimum refractive index of the waveguide needs to be equal or larger than unity ($n_{min} \geq 1$). Therefore, the constant $c_1$ is determined in such a way as to achieve $n_{min}=1$. The maximum value of the refractive index distribution achieved by CM may be too large to be fabricated by using typical low-loss dielectric materials such as silica ($SiO_2$). Therefore, the resultant refractive index distribution may need to be modified in order to confine it below the maximum refractive index of the utilized dielectric material at the design wavelength ($\lambda_d$). An important factor in designing the waveguides with high performance is to minimize the reflection on the input and output boundaries of the device ($\Gamma_1$ and $\Gamma_3$), which in the case of dielectric materials can be done by index-matching to the surrounding environment (i.e. vacuum). The modification of the refractive index distribution as well as the index-matching are performed according to the method proposed in [13], which can be used to decrease the maximum value of the refractive index and apply index-matching to the boundaries, while keeping the propagation path of the EM filed in the waveguide unchanged. The sizes of the waveguide are also of great importance; firstly, by applying CM, the optical size of the device for the longest operating wavelength ($\lambda_\infty$) needs to be in the range of geometric optics (i.e. $c_1/\lambda_\infty \gg 1$), and secondly by applying the modification method, the modification coefficients also need to be taken into account [13].

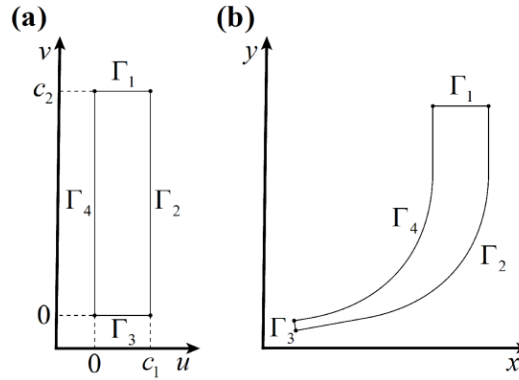

**Figure 2. (a)** Shows the virtual domain. **(b)** Shows the physical domain.

The gradient refractive index derived by CM design can be realized by non-resonant bulk metamaterials [26]. Optical metamaterials are the artificial materials consist of sub-wavelength building blocks which enable unusual interactions with the EM filed. In the case of non-resonant bulk metamaterials, the size of the building blocks are considerably smaller than the minimum wavelength of interest, which eliminates the possibility of resonance of the building blocks with the EM field. Therefore, for this type of metamaterials an effective refractive index can be defined, which is the mixture of the refractive indices of the materials of the building blocks [26]. By considering the building blocks composed of silica and vacuum, the effective index is a function of the filling factor ($R$), which is the ratio between the volume of the silica and the total volume of the building block. Therefore, the refractive index can be tuned by controlling the filling factor, and the gradient index of the waveguide can be realized by gradually varying the filling factor from zero to unity ($0<R<1$).

The refractive index distribution achieved by CM technique is valid merely at the design wavelength $\lambda_d$. In order to achieve the index distribution for other wavelengths, the dispersion curve of the effective medium needs to be obtained, which in this work is achieved through a modified S-parameter retrieval technique [27], for different filling factors over the desired wavelength range. By considering the dispersion curve of the effective medium, it is apparent that the index distribution of the waveguide varies over different wavelengths. Therefore, the refractive index distribution needs to be designed to have a good performance over the broad spectrum of solar radiation. In this study, a wavelength range of $\lambda_0=200$nm to $\lambda_\infty=4500$nm is considered, which according to the spectral solar luminosity $L_\odot(\lambda)$ given by the black body radiation at the effective solar temperature of $T_\odot=5772K$ [4], covers about 99% of the total solar radiation power. Since the efficiency of the waveguide declines for the wavelengths distant from $\lambda_d$, the design wavelength needs to be carefully chosen to minimize this adverse effect. In order to decrease the total undesirable effect of the broadband response of the waveguide for the wavelengths distant from $\lambda_d$, the design wavelength is simply considered to be the center of $n_{SiO2}(\lambda).L_\odot(\lambda)$ surface as

$$\lambda_d = \frac{\int_{\lambda_0}^{\lambda_\infty} \lambda n_{SiO2} L_\odot \, d\lambda}{\int_{\lambda_0}^{\lambda_\infty} n_{SiO2} L_\odot} \tag{4}$$

where $n_{SiO2}(\lambda)$ is the dispersion curve of silica.

Since the solar radiation is non-polarized, the bulk metamaterials used to realize the graded index waveguide need to be polarization-invariant and isotropic. The structures with 3D topology are good candidates to realize a three-dimensional isotropic polarization-invariant structure with a two-dimensional index variation (in $x_w y_w$ plane). Such structures can be realized by a lattice composed of connected network of dielectric spheres in air, or air spheres in dielectric, as well as the rod-connected or wood-pile structures, which among them, the rod-connected and wood-pile structures can cover the whole range of filling factor ($0 \leq R \leq 1$) [28]. These structures have been utilized to fabricate 3D gradient index devices, including wave concentrator [29], and Eaton lens [30], which operate in microwave regime, spherical Luneburg lenses for terahertz frequencies [31], and optical wavelengths [32], as well as three-dimensional invisibility cloak for optical wavelengths [33]. The nanostructures required to realize the gradient-index three-dimensional structures for optical frequencies can be fabricated by direct laser writing which an ultrafast laser polymerizes a photosensitive material via two- or multiphoton absorption [32, 33].

**Estimation of the sail's performance**
In order to estimate the performance of the proposed gradient refractive sail under the incoherent solar radiation, the ray tracing technique is utilized. By having the power and direction of the input and output rays, the radiation pressure can be given as [4]

$$\boldsymbol{P} = \frac{1}{c} \int_{\lambda_0}^{\lambda_\infty} W_T \left( \sum_{N_i=1}^{N_i=N_{i\max}} R_i^{N_i} \boldsymbol{k}_i^{N_i} - \sum_{N_o=1}^{N_o=N_{o\max}} R_o^{N_o} \boldsymbol{k}_o^{N_o} \right) d\lambda \tag{5}$$

where $N_i$ and $N_o$ are the index of summation regarding the input and output beams, with the maximum $N_{i\max}$ and $N_{o\max}$ number of input and output beams, respectively. The nomenclature $W_T$ is the spectral solar power density at the distance $r_\odot$ from the Sun which can be given by having the spectral solar luminosity $L_\odot$ as

$$W_T = \frac{L_\odot}{4\pi r_\odot^2} \tag{6}$$

The unit vectors $\boldsymbol{k}_i$ and $\boldsymbol{k}_o$ define the direction of a single input and output ray, respectively. The coefficients $R_i$ and $R_o$ are the relative power density of each input and output ray, respectively, to the spectral solar power density at the wavelength of the ray, and can be given as

$$R_o^{N_o} = \frac{W_o^{N_o}}{W_T}, \quad R_i^{N_i} = \frac{W_i^{N_i}}{W_T} \tag{7}$$

where $W_i$ and $W_o$ are the spectral power density of a single input and output beam, respectively. By considering a number of $N_{i\max}$ identical input rays, the spectral power density of a single input ray can be given as

$$W_i^{N_i} = \frac{W_T}{N_{i\max}} \tag{8}$$

The radiation pressure vector $\boldsymbol{P}$ given by the components $P_x$ and $P_y$ in the waveguide-fixed frame shown in Fig. 1a, can be computed by the results of ray tracing analysis for any given incidence angle. However, in this work a single normal incidence angle is considered, which can greatly simplify the orbital control of the sail, as we will show next.

## Orbital dynamics and control

The sails capable of generating a tangential radiation pressure such as the gradient-index sail, refractive sail [4], diffractive sail [6], and anomalous reflective sail [7], can provide the possibility of orbital control through simple attitude maneuvers, by keeping a sun-pointing attitude and controlling the attitude merely around the sun-sail axis.

The orbital dynamics of the sail is given in an orbital reference frame in the heliocentric inertia system shown in Fig. 3, where the unit vectors $\hat{r}$, $\hat{h}$, and $\hat{t}$ are along the radial, orbit normal and transverse directions, respectively. In the case of conventional reflective sails, the radiation pressure vector is defined by clock and cone angles [18]. However, for tangential-radiation-pressure-generating sails, the attitude of the sail is passively stabilized at a sun-pointing attitude, and the orientation of the tangential radiation pressure is given by the angle $\delta$, as shown in Fig. 3. Here we consider the radiation pressure vector at the normal incidence angle at 1 AU distance from the sun, described by the normal and tangential components $P_n$ and $P_p$, respectively, given in orbital frame shown in Fig. 3. The components $P_n$ and $P_p$ are identical to the $P_y$ and $P_x$ of the waveguide-fixed frame, respectively, which are computed by Eq. 5 using ray tracing technique. The clock angle $\delta$ is defined as the angle between the orbital angular momentum $\hat{h}$ and $P_p$.

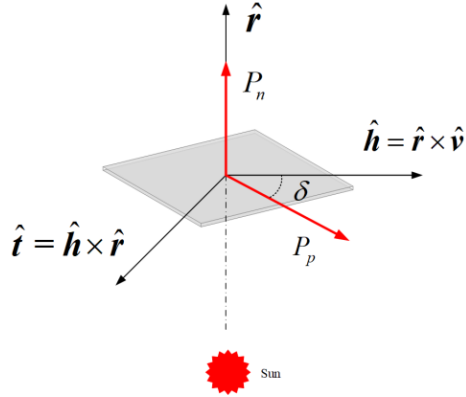

**Figure 3.** Definition of the reference frame for orbital dynamics.

The characteristic acceleration $a_c$ is given as the total radiation pressure acceleration of the sail experienced at a sun-pointing attitude and a distance of 1 astronomical unit (AU) from the sun [18]. This is similar to the definition of the characteristic acceleration definition for a conventional perfectly reflective solar sail, except that the characteristic acceleration for a gradient-index sail refers to the magnitude of the vector sum of the accelerations in two perpendicular directions. For a given area-to-mass ratio $\varepsilon$, given in $m^2/kg$, the characteristic acceleration which is the acceleration of the sail caused by solar radiation pressure at 1 AU, can be written as

$$a_c = \varepsilon\sqrt{P_n^2 + P_p^2} \tag{9}$$

The normal and tangential radiation acceleration of the gradient-index sail at the distance of $R$ (AU) from the sun can be expressed as

$$\begin{cases} a_n = \sigma_n \dfrac{a_c}{R^2} \\ a_p = \sigma_p \dfrac{a_c}{R^2} \end{cases} \tag{10}$$

where the coefficients $\sigma_n$ and $\sigma_p$ are given as $\sigma_n = P_n/\sqrt{P_n^2 + P_p^2}$ and $\sigma_p = P_p/\sqrt{P_n^2 + P_p^2}$. Thus, the sail's total acceleration vector can be written as

$$\boldsymbol{a}_s = \frac{a_c}{R^2}\left[\sigma_n\hat{\boldsymbol{r}} + \sigma_p\left(\cos\delta\hat{\boldsymbol{h}} + \sin\delta\hat{\boldsymbol{t}}\right)\right] \tag{11}$$

By ignoring the perturbation forces except for the solar gravitation and solar radiation pressure force, the dynamic equations of the sail in the heliocentric inertial system can be given as

$$\begin{cases} \dot{\boldsymbol{R}} = \boldsymbol{V} \\ \dot{\boldsymbol{V}} = -\dfrac{\boldsymbol{R}}{R^3} + \boldsymbol{a}_s \end{cases} \tag{12}$$

where $\boldsymbol{R}$ and $\boldsymbol{V}$ are the position and velocity vectors of the sail (known as state vectors) in the heliocentric inertial system. For the convenience of calculations, the length unit is normalized with the astronomical unit and the time unit is normalized with $1/2\pi$ year [21]. Therefore, $\boldsymbol{R}$ and $\boldsymbol{V}$ are presented in AU and AU/year units, respectively.

The time-optimal trajectories in interplanetary mission using conventional reflective sail have been deeply investigated [18, 19, 20, 21]. However, as we will show, the time-optimal control strategy for interplanetary transfer missions regarding the tangential-radiation-pressure-generating sails can be applied in a simpler manner. By considering the dynamic equation given in Eq.12, the optimization goal is to minimize the transfer time subject to the boundary conditions as interplanetary transfer missions. By considering the co-state vectors for position and velocity of the sail as $\lambda_R$ and $\lambda_V$, the Hamiltonian function can be defined as

$$H = -1 + \lambda_R \cdot V + \lambda_V \cdot \left( -\frac{1}{R^3} R + \frac{1}{R^2} \left( \sigma_n \hat{r} + \sigma_p \hat{p} \right) \right) \tag{13}$$

and the Euler-Lagrange equations is given as

$$\begin{cases} \dot{\lambda}_R = -\frac{\partial H}{\partial R} = \frac{1-\sigma_n}{R^3} \lambda_V - \frac{(3-3\sigma_n)(\lambda_V \cdot R) + 2\sigma_p R(\lambda_V \cdot \hat{p})}{R^5} R \\ \dot{\lambda}_V = -\frac{\partial H}{\partial V} = -\lambda_R \end{cases} \tag{14}$$

where the unit vector $\hat{p}$ represents the direction of the tangential radiation acceleration as

$$\hat{p} = \cos\delta \hat{h} + \sin\delta \hat{t} \tag{15}$$

Since the sail keeps a stable sun-pointing attitude at all time, the coefficient $\sigma_n$ as given in Eq.13, is not tunable. Therefore, According to Eq.13, the value of the Hamiltonian function can be altered merely by tuning the direction of unit vector $\hat{p}$, which according to Eq. 15 can be applied by controlling he angle $\delta$. In order to maximize the Hamiltonian function, the optimal angle $\delta^*$ needs to maximize the projection of $\hat{p}$ in the direction of $\lambda_V$. The direction of $\lambda_V$ can be given as a unit vector by cone and clock angles of $\tilde{\alpha}$ and $\tilde{\delta}$, respectively, as

$$\hat{\lambda}_V = \cos\tilde{\alpha}\hat{r} + \sin\tilde{\alpha}\cos\tilde{\delta}\hat{h} + \sin\tilde{\alpha}\sin\tilde{\delta}\hat{t} \tag{16}$$

Consequently, according to Eq.15 the optimal control law can be obtained as

$$\delta^* = \tilde{\delta} \tag{17}$$

The optimal control problem is solved by the numerical indirect shooting method [34]. By guessing the initial co-state vectors and the departure and arrival times, we obtain the arrival states and co-states of the sail by integrating the Eqs. 12 and 14, which will be used in an iteration to check the optimality conditions.

## Results and Discussion

In this section we present the results of the waveguide design given as refractive index distribution, as well as the results of the ray tracing and the measured total radiation pressure. We will then show the feasibility of the gradient-index sail in classical trajectory transfer optimal control problem by two numerical examples.

### Refractive index distribution

The sizes of the waveguide are defined according to the longest wavelength of light considered in this work and the method given in [13]. The sizes of the waveguide are given as follows; $\alpha_w$=80 deg, $r_w$=20 μm, $w_w$=6.15 μm, $l_{wi}$=8 μm, $l_{wo}$=8 μm. By considering the wavelength range considered in this work ($\lambda_0$=200nm and $\lambda_\infty$=4500nm), the design wavelength is computed by Eq. 4 to be $\lambda_d$ =876nm. The resultant refractive index distribution using CM method is shown in Fig. 4a. It can be seen that the refractive index has a minimum value of $n_{min}$=1, expectedly. The maximum refractive index, however, has a very large value of $n_{max}$=6.94, which can't be realized by silica. The refractive index is then modified by the method given in [13], and shown in Fig. 4b. It can be seen that the modified index has a maximum value of $n_{max}$=1.45 given for the design wavelength $\lambda_d$ =876nm. Since the modified $n_{max}$ is smaller than the refractive index of silica at this wavelength, given as 1.4521 [35], the index distribution can be realized by tuning the filling factor over the surface of the waveguide.

The refractive index distribution shown in Fig. 4b, which is given for the wavelength $\lambda_d$, results in exact guiding of the EM field through the designed path of the waveguide. However, the index distribution which is realized as the effective index of a bulk metamaterial by altering the filling factor, varies over the range of the operating wavelengths, according to the dispersion curve of the bulk metamaterial. The dispersion curve of the bulk metamaterial made of the building blocks composed of silica and air, is obtained through a modified S-parameter retrieval technique [27], for different filling factors over the desired wavelength range. By having the dispersion relationship of the bulk metamaterial, the refractive index distribution can be derived for the wavelengths other than $\lambda_d$ =876nm. As an example, the extended refractive index distributions for four wavelengths of $\lambda$=200nm, $\lambda$=876nm, $\lambda$=3000nm, and $\lambda$=4500nm are shown in Fig. 5. It can be seen that the refractive index decreases for longer wavelengths due to the dispersion curve.

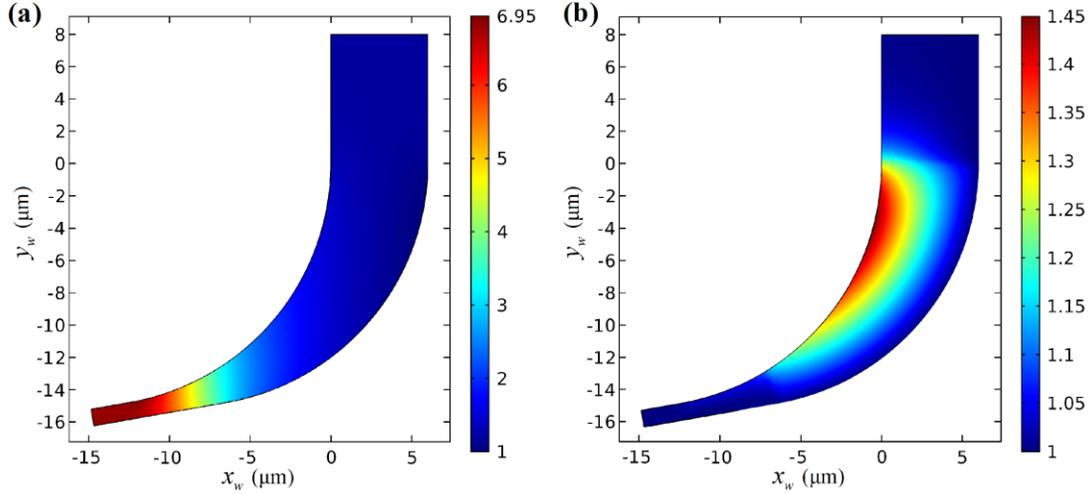

**Figure 4.** (**a**) The refractive index distribution of the waveguide generated by CM. (**b**) The modified index.

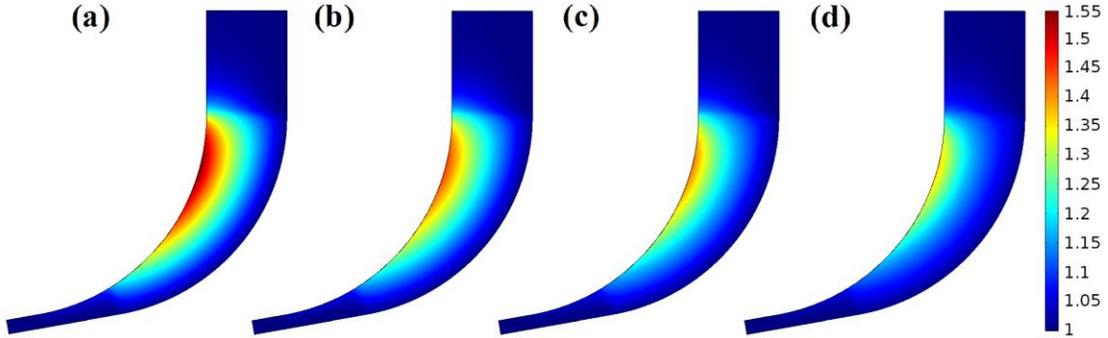

**Figure 5.** The refractive index distribution of the waveguide at different wavelengths. (**a**) $\lambda$=200nm. (**b**) $\lambda$=876nm. (**c**) $\lambda$=3000nm. (**d**) $\lambda$=4500nm.

**Ray tracing results**
By considering the direction of the sail relative to the waveguide-fixed frame shown in Fig. 1a, the unit vector of the input rays is given as $\boldsymbol{k}_i$=[0 -1 0], and the output unit vectors is computed by ray tracing technique. The ray tracing is performed by considering the sail at a distance of 1 AU from the Sun. The input and output beams are modeled with a maximum number of $N_{imax}$=100 and $N_{omax}$=2100, input and output rays, respectively. The spectral power density of the input rays are given by the black body radiation at the temperature of $T_\odot$=5772K. The results of the ray tracing over several wavelengths are shown in Fig. 6, where the color of each ray shows the relative spectral power density of each ray ($R_i$ and $R_o$). It can be seen that at the wavelength $\lambda$=876nm (Fig. 6d), which is the design wavelength of the waveguide, the sail can accurately guide the beams through the designated path. However, by getting farther from the design wavelength, the ability of the sail in directing the rays is decreased, which results in a lower performance over longer and shorter wavelengths.

The absolute values of the term $\sum_{N_i=1}^{N_i=100} R_i^{N_i} \boldsymbol{k}_i^{N_i} - \sum_{N_o=1}^{N_o=2100} R_o^{N_o} \boldsymbol{k}_o^{N_o}$ from Eq. 5 are computed by using the results of the ray tracing for 62 different wavelengths over $\lambda_0$=200nm to $\lambda_\infty$=4500nm, for the unit vectors along $x_w$ and $y_w$, separately, which are shown in Fig. 7. The total radiation pressure along $x_w$ and $y_w$ is computed by the integration over the wavelength range, as presented in Eq. 5. For the design parameters presented in this paper, the total radiation pressure along $x_w$ and $y_w$ axes are computed as $P_x$ = 4.126×10$^{-6}$ (N/m$^2$) and $P_y$ = -3.346×10$^{-6}$ (N/m$^2$). By considering the maximum possible theoretical value of tangential radiation pressure at the Earth orbit as $P_{max}$ = 4.557×10$^{-6}$ Pa [12], the generated tangential radiation pressure by gradient-index sail ($P_x$) is 90.5 % of the maximum possible radiation pressure ($P_{max}$). By considering a near-circular sun-centered orbit which is the case for interplanetary missions, and by keeping a constant sun-pointing attitude, nearly the whole $P_x$ can be pointed towards the orbital velocity. Maximizing the radiation pressure along the orbital velocity is the local optimal condition for increasing the orbital energy and changing the orbital radius. In the case of an ideal conventional reflective sail, the maximum possible radiation pressure along the orbital velocity is 76.98 % of $P_{max}$ at an angle of 35.26 degrees between the sail's normal and the sun-sail vectors. Therefore, the magnitude of the useful radiation pressure for local optimal orbital control of a gradient refractive sail is larger than that of an ideal conventional sail. This higher radiation pressure efficiency along with the simple orbital control realized by a sun-pointing attitude as discussed before, are the most important advantages of the gradient-index refractive sail over conventional sails.

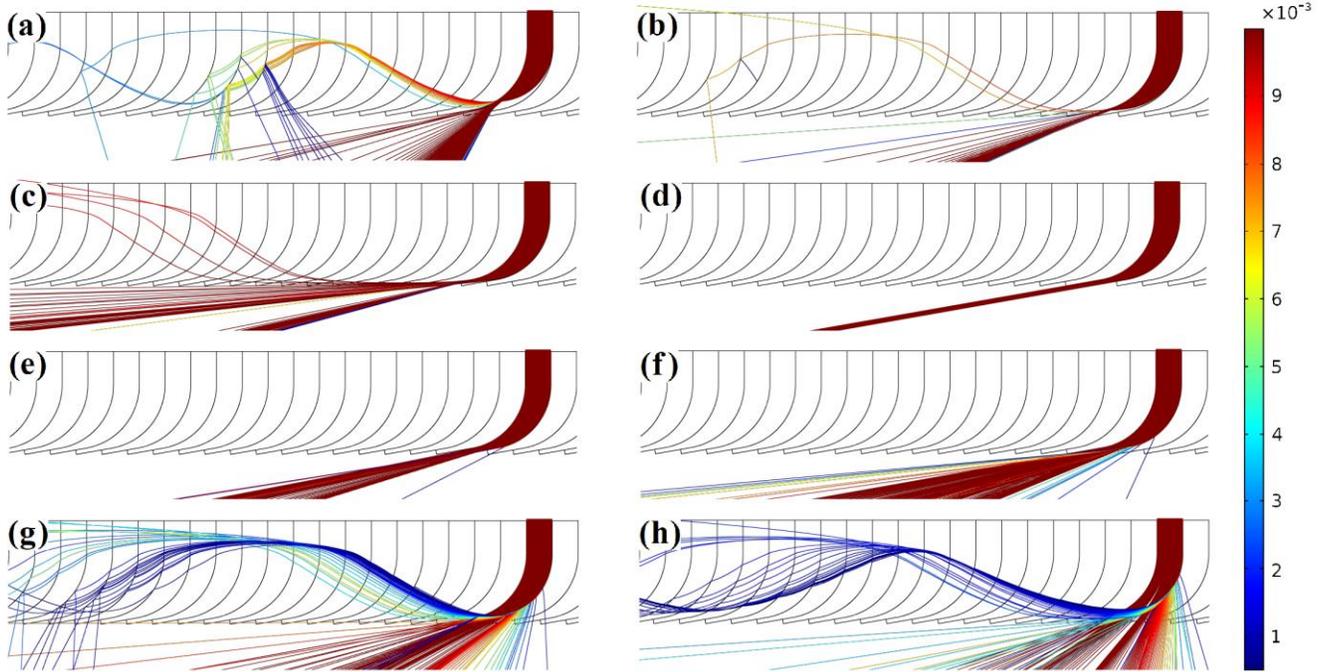

**Figure 6.** The results of the ray tracing. **(a)** λ=200nm. **(b)** λ=400nm. **(c)** λ=600nm. **(d)** λ=876nm. **(e)** λ=1115nm. **(f)** λ=1515nm. **(g)** λ=2500nm. **(h)** λ=4500nm.

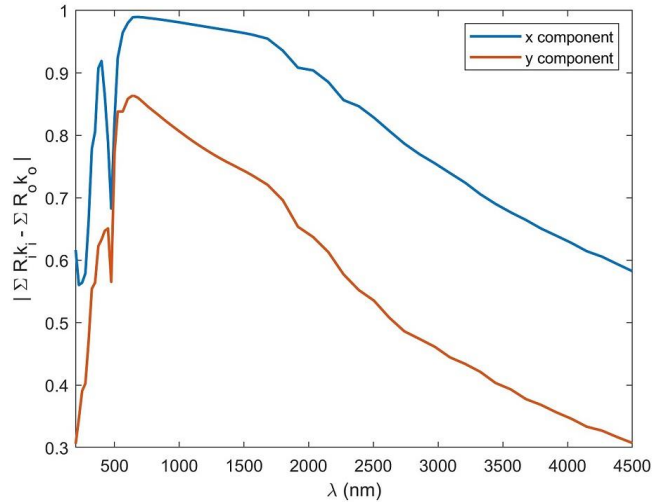

**Figure 7**. The term $|\Sigma R_i \mathbf{k}_i - \Sigma R_o \mathbf{k}_o|$ computed for $x$ and $y$ axes separately, using the ray tracing results.

### Optimal orbital control

We show the feasibility of the proposed orbital control in classical trajectory transfer optimal control problem, by using two examples. According to the sizes of the waveguide, a gradient-index sail with a total area-to-mass ratio of $\varepsilon = 33 \text{ m}^2/\text{kg}$ and accordingly a characteristic acceleration of $a_c = 0.175 \text{ mm/s}^2$ is chosen. In the first example the classical time-optimal transfer mission from the Earth to Venus is considered. The solution of angle $\delta$ is plotted in Fig. 8a, and the transfer trajectory along with the tangential radiation pressure vectors are shown in Fig. 8b. The launch window is set between the beginning of 2010 and the end of 2022. As shown in Fig. 8b, within this launch window, the optimal launch and rendezvous dates are obtained as 29 April 2021 and 29 July 2022, respectively, which correspond to a flight time of 456.3 days.

In order to show the feasibility of the proposed orbital control in time-optimal transfer missions between two bodies with different semi-long axis, eccentricity, and orbital inclination, an example mission from Earth to asteroid (433) Eros is considered. The area-mass ratio and characteristic acceleration are the same with the first example. The solution of angle $\delta$ is shown in Fig. 9a, and the 2D and 3D transfer trajectories along with the tangential radiation pressure vectors are shown in Fig. 9b and Fig. 9c, respectively. As shown in Fig. 9b and Fig. 9c, the gradient-index sail departs the Earth on 11

November 2024 and flies for 1207.3 days before arrival to Eros on 2 March 2028. The results of these analyses show that the proposed orbital control strategy is able to complete the interplanetary transfer missions by only controlling the direction of the tangential radiation pressure vector.

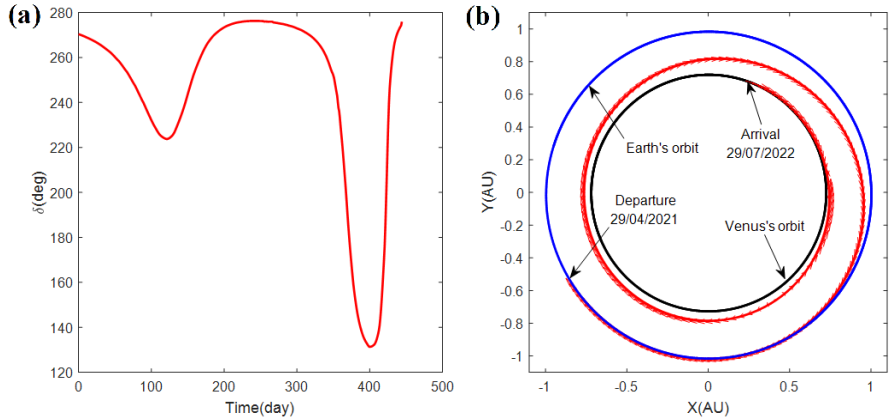

**Figure 8**. Optimal transfer solution from Earth to Venus. **(a)** Control profile of angle δ. **(b)** Transfer trajectory.

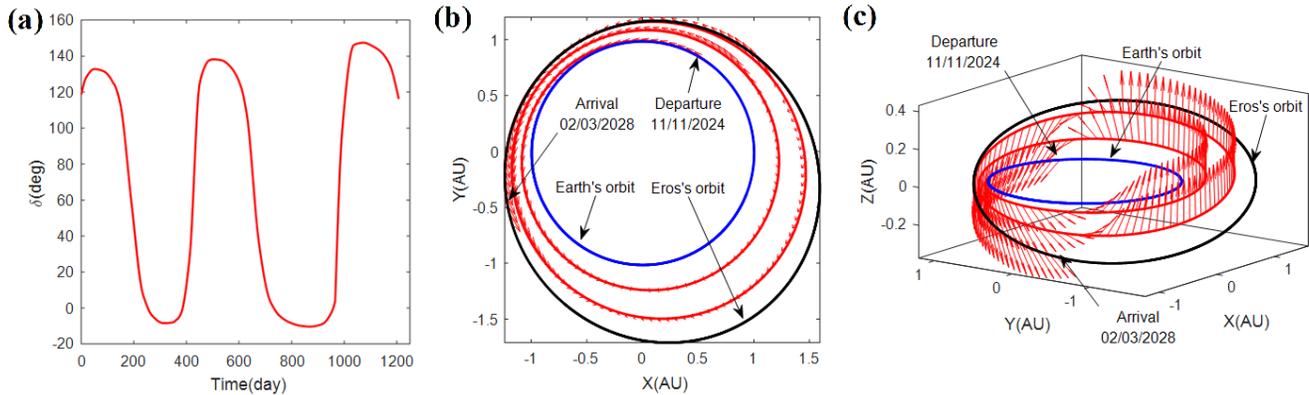

**Figure 9**. Optimal transfer solution from Earth to Eros. **(a)** Control profile of angle δ. **(b)** Transfer trajectory in 2D view. **(c)** Transfer trajectory in 3D view.

## Conclusion

We have found that solar sails composed of an array of waveguides with gradient-index medium can achieve a large efficiency of 90.5 % in generating the tangential radiation pressure at a normal incidence angle. The proposed gradient-index sail can be regarded as the first step in designing high efficiency broadband photonic propulsion devices by employing graded index media for generating radiation pressure under the broad spectrum of solar radiation. It was shown how the refractive index distribution designed by conformal mapping for a single wavelength can be extended to a wide range of wavelengths, which enables the utilization of ray tracing technique for estimation of the magnitude and direction of a broadband radiation pressure applied to gradient-index media. It was also found that the optimal orbital control of tangential-radiation-pressure-generating sails at normal incidence angle, can be simply applied by merely controlling the attitude of the sail around the sail's normal axis, while keeping a passively stable sun-pointing attitude. We have shown that by controlling this single angle, the orbital axes, eccentricity, and inclination can be tuned. By considering the large tangential radiation pressure applied to a gradient-index sail, applying such a control method can provide simplicity and efficiency in the same time. Gradient-index films can also provide an attitude control mechanism for controlling the attitude of any sail type along the normal axis at a normal incidence angle, in a similar manner to refractive films.

## Acknowledgements

S.F. acknowledges the support from China Scholarship Council (CSC) for funding his PhD studies.

## Author contributions statement

S.F. developed the concept and theoretical model of gradient-index sail, designed and simulated the sail, and wrote and polished the manuscript. Y.S. developed the optimal control model, simulated the time-optimal transfer missions, and wrote the manuscript. S.G. provided the intellectual input, coordinated and supervised the project. All authors participated in the discussion, interpreted the data, and commented on the manuscript.

## Additional Information

**Competing Interests:** The authors declare no competing interests.